\documentclass[sigconf]{acmart}

\usepackage{color}
\usepackage{graphicx}
\usepackage{subfigure} 
\usepackage{booktabs}
\usepackage{multirow}
\usepackage{placeins}
\usepackage{float}

\AtBeginDocument{%
  \providecommand\BibTeX{{%
    \normalfont B\kern-0.5em{\scshape i\kern-0.25em b}\kern-0.8em\TeX}}}

\copyrightyear{2023}
\acmYear{2023}
\setcopyright{acmlicensed}\acmConference[CHI '23]{Proceedings of the 2023 CHI Conference on Human Factors in Computing Systems}{April 23--28, 2023}{Hamburg, Germany}
\acmBooktitle{Proceedings of the 2023 CHI Conference on Human Factors in Computing Systems (CHI '23), April 23--28, 2023, Hamburg, Germany}
\acmPrice{15.00}
\acmDOI{10.1145/3544548.3580948}
\acmISBN{978-1-4503-9421-5/23/04}

\begin{document}

\title{PopBlends: Strategies for Conceptual Blending with Large Language Models}

\author{Sitong Wang}
\affiliation{
  \institution{Columbia University}
  \city{New York}
  \state{NY}
  \country{USA}
}
\email{sw3504@columbia.edu}

\author{Savvas Petridis}
\affiliation{
  \institution{Columbia University}
  \city{New York}
  \state{NY}
  \country{USA}
}
\email{sdp2137@columbia.edu}

\author{Taeahn Kwon}
\affiliation{
  \institution{Columbia University}
  \city{New York}
  \state{NY}
  \country{USA}
}
\email{taeahn.kwon@columbia.edu}

\author{Xiaojuan Ma}
\affiliation{
  \institution{Hong Kong University of Science and Technology}
  \city{Hong Kong}
  \country{China}
}
\email{mxj@cse.ust.hk}

\author{Lydia B. Chilton}
\affiliation{
  \institution{Columbia University}
  \city{New York}
  \state{NY}
  \country{USA}
}
\email{chilton@cs.columbia.edu}

\renewcommand{\shortauthors}{Wang, et al.}

\begin{abstract}
Pop culture is an important aspect of communication. On social media people often post pop culture reference images that connect an event, product or other entity to a pop culture domain. Creating these images is a creative challenge that requires finding a conceptual connection between the users' topic and a pop culture domain. In cognitive theory, this task is called conceptual blending. We present a system called PopBlends that automatically suggests conceptual blends. The system explores three approaches that involve both traditional knowledge extraction methods and large language models. Our annotation study shows that all three methods provide connections with similar accuracy, but with very different characteristics. Our user study shows that people found twice as many blend suggestions as they did without the system, and with half the mental demand. We discuss the advantages of combining large language models with knowledge bases for supporting divergent and convergent thinking.
\end{abstract}



\begin{teaserfigure}
\centering
\includegraphics[width=1\textwidth]{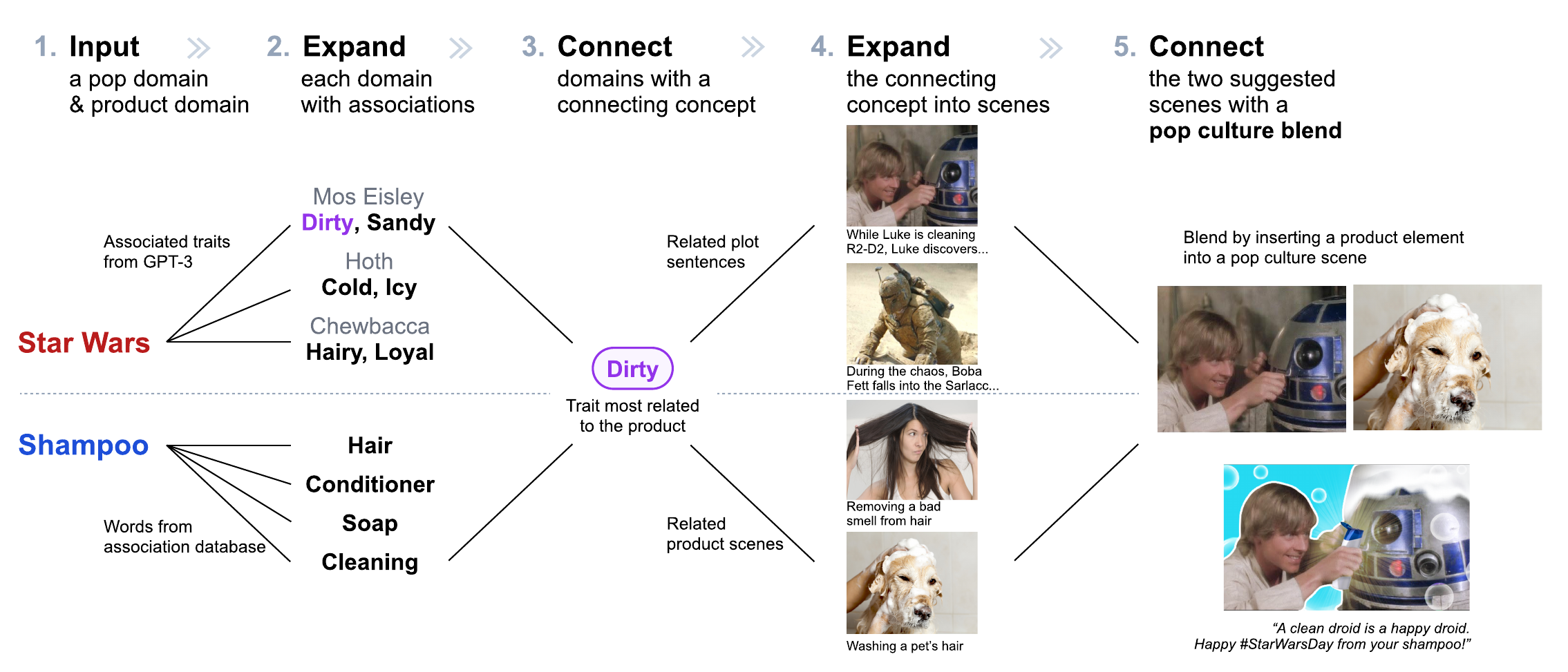}
\caption{An example of the PopBlends system automatically suggesting pop culture blends for the inputs of Star Wars and shampoo. The system first expands both inputs into associations, then finds connections between the associations. For the best connections, the system searches for images of scenes that are related to the inputs (Star Wars-related images \textcopyright Lucasfilm Ltd.). We show an artist rendering of one of the blend suggestions.}
\label{fig:system_teaser}
\end{teaserfigure}

\maketitle

\section{Introduction}
\begin{figure*}
\centering
\includegraphics[width=0.9\textwidth]{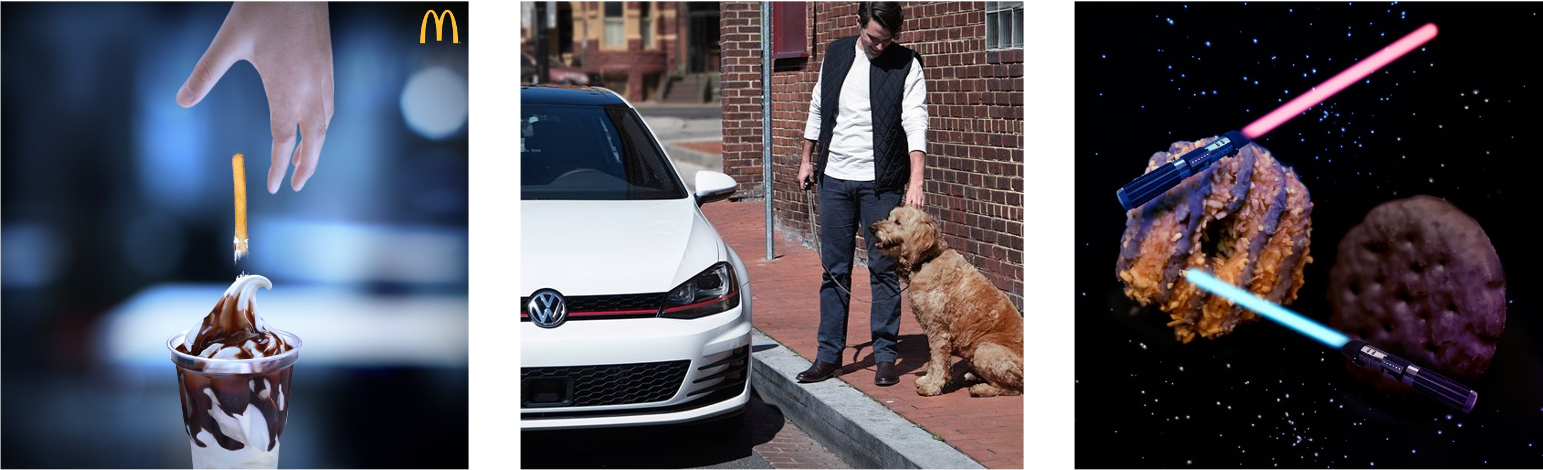}
\caption{Pop culture blends for Star Wars Day collected on Twitter from (a) McDonald's, (b) Volkswagen, and (c) Girl Scouts.}
\label{fig:formative_study_samples}
\end{figure*}

Pop culture is an important aspect of communication. 
References to memorable moments in television, film, and other mediums pervade Internet communication as well as everyday conversation. 
One particularly creative type of messaging used on social media is to blend an organization’s product or service with images from a trending pop culture domain. 
For example, on Star Wars Day (May the Fourth) many organizations blend their brand or product with images from Star Wars to post on social media. 
Examples include someone using the force to extract a french fry from a McDonald's ice cream, Han and Chewbacca getting into their Volkswagen, and light and dark Girl Scout Cookies having a lightsaber battle (See Figure \ref{fig:formative_study_samples}).
These images are  helpful for online campaigns because they capture attention and connect the product to something people already know and like---such as the characters of a popular film or TV show ~\cite{tweetorials}. 
We call such images \textit{pop culture blends} because they blend a product or service with references to a pop culture domain.

Underlying the creation of a pop culture blend is the cognitive task of conceptual blending ~\cite{FAUCONNIER1998133}.
Conceptual blending is a type of combinatorial creativity ~\cite{BODEN1998347} that connects two different input domains into a novel outcome through a \textit{connecting concept}. 
Conceptual blending ultimately requires divergent and convergent thinking processes ~\cite{tversky_creativity}. 
Models of conceptual blending lead us to expect the involvement of two divergent and convergent processes: one to identify the connecting concept between the two domains and the other to find the context and images to implement the connecting concept into a blend.  
Figure \ref{fig:system_teaser} shows an example of this process for Star Wars and shampoo, where Star Wars and shampoo are both expanded until a connecting concept can be found, in this case, ``dirty''. 
Then the concept of ``dirty'' was expanded for Star Wars and shampoo, respectively, until a scene image can be found from each domain to be blended, such as a scene of Luke cleaning R2-D2 with shampoo suds inserted.

In general, the conceptual blending process is difficult because the divergent process requires recalling multiple, diverse pieces of information and the convergent process requires a polynomial search over the information to find pairs that can be blended.
Typically, this search process is time-consuming for people because the design space is vast and the useful connections are rare. 
Although computers possess the computational power necessary for the job, they tend not to have the cultural knowledge, associations, and commonsense required for this task. 
Divergent thinking requires finding associations that are typically not stored in structured databases such as linking shampoo to a suds, or Star Wars to the scene of ``Luke cleaning R2-D2''. 
Crowdsourced association databases such as Small World of Words ~\cite{swow} have some coverage of human associations but often lack pop culture knowledge. 
Trained word embeddings such as GloVe \cite{glove} and word2vec \cite{word2vec} often lack associative information, or contain just as much noise as signal~\cite{swow_better_w2v}. 
Additionally, convergent thinking requires commonsense reasoning to find logical connections between the associations. Attempts to catalog commonsense have found it to be an overwhelmingly large task ~\cite{Lenat1995Cyc} and they thus have focused on particular types of relationships like if-then ~\cite{choi_atomic}, which are impressive but not broad enough for most creative tasks.

Recent advances in the generative ability of large language models are poised to create a paradigm shift for computational creativity. While previous pretrained language models like GPT~\cite{gpt1} and GPT-2~\cite{gpt2} generated fluent text, they did not possess human-like abilities to reason. 
However, GPT-3 \cite{gpt}, which has billions more parameters than GPT-2, can do tasks that involve associative knowledge and commensense reasoning---tasks that previously required human intelligence. 
Because large language models have been trained on Internet-scale data, they have information related to most topics people talk about, including pop culture~\cite{gpt, gpt_as_kb}. 
For example, GPT-3 can be prompted for associations by asking:``List 5 adjectives you associate with Chewbacca from Star Wars'', and GPT-3 will respond ``brave, loyal, gentle, hairy, heroic''.
Moreover,  GPT-3 has the ability to perform sentence completion and zero-shot learning, meaning it can be prompted with text like ``Which character in Star Wars is most related to shampoo? Why?'' and it will return a character and a reason such as ``Princess Leia, because she is known for her iconic hairstyles.'' 
However, GPT-3's answers are not always accurate or useful, thus the challenge lies in how to build system architectures around GPT-3 that people will find useful and powerful. 

We present a system called PopBlends that automatically performs divergent and convergent steps to suggest pop culture blends.  
First, the user inputs two concepts---a pop culture domain such as Star Wars, and a product, service, or other entity such as shampoo. 
The system starts by performing a divergent and convergent step to find multiple connecting concepts with three different strategies using various levels of support from the large language model GPT-3. 
Based on each connecting concept, the system conducts a second divergent and convergent process to find the scenes that bridge the two inputs through the connecting concepts.
The system then finds images related to each scene and the user can select the images to combine into a pop culture blend. 
Ultimately, the aim is to help amateur designers ideate connections between a product or service and a pop culture domain. 

This paper makes the following contributions:
\begin{itemize}\itemsep0em
\item A comparison of three GPT-3 strategies to find connecting concepts: No-GPT (traditional NLP and knowledge extraction only), Half-GPT (combining knowledge extraction and GPT), and Full-GPT (relying solely on GPT for connections). 
We find they are all effective and produce surprisingly different results.
\item PopBlends, a system that suggests image pairs to amateur designers to help them make a pop culture blend. 
The system uses two rounds of divergent and convergent processes: one to find connecting concepts between the pop culture and product domains and another to find scenes and images related to the connecting concept that can be blended. 
\item An annotation study demonstrating the usefulness of GPT-3 to list rich pop culture associations and perform commonsense reasoning over those associations.
\item A user study showing that the PopBlends system helped people come up with twice as many pop culture blends, and with half of the mental demand compared to a baseline condition of brainstorming and Internet search.
\end{itemize}
We conclude our work with a discussion of the advantages of combining large language models with knowledge bases for supporting divergent and convergent thinking.
\enlargethispage{12pt}
\section{Related Work}
\subsection{Conceptual Blending and Pop Culture Blends}
Conceptual blending has been shown to be a beneficial part of the creative design process~\cite{dow_sharing_designs}. 
Conceptual blending or conceptual integration is a cognitive science theory for how to combine elements of two familiar inputs or concepts to make new meaning or novel artifacts \cite{FAUCONNIER1998133}. 
The theory describes the blending process in two steps: first finding a concept that connects the two inputs, then based on the connecting concept, finding which elements from each input can be mapped to each other and blended into a new artifact or idea---like a houseboat being a combination of a house and a boat, but retaining the essential properties of each.
It is a type of creativity that Boden~\cite{BODEN1998347} categorizes as combinational. 
It is related to other creative problems such as constructing metaphors, innovation by analogical reasoning, and abstract problem solving \cite{Veale2019}.

Typical computational approaches to conceptual blending are to expand or generalize one or more of the inputs, then search over the expansion to find connections ~\cite{framework_comp_blending}.
This is closely related to divergent and convergent thinking---the cognitive underpinnings of creativity~\cite{tversky_creativity}.
These two steps (i.e., expansion and search) are difficult tasks to be done by humans alone or to be fully automated.
Expanding or generalizing an input is difficult because it is ``information hungry'' ~\cite{Veale2019}. 
Classic approaches that follow the structure-mapping approach require hand-crafted datasets ~\cite{Falkenhainer89thestructuremapping, computation_and_blending, Pereira_conceptual_blending}, which are expensive and difficult to scale. Internet search has been proposed as a potential solution to the information hungry problem, but often it is hard to answer simple, commonsense, and associative reasoning from Internet searches.
Searching over the expansions to find connections is challenging due to the polynomial number of possible combinations. 
Moreover, most combinations are not meaningful – randomly combining parts of a house and a boat is not likely to result in something functional. 
Although this can be seen as a search and optimization problem~\cite{Pereira_conceptual_blending}, it is difficult to know what makes a good connection and thus hard to computationally search and optimize for.
Thus, a good approach is to combine people’s abilities for judgment with the machine’s abilities to search.

The cognitive task of conceptual blending is the basis of creating pop culture blends. 
Previous research on other conceptual blending tasks ~\cite{metamap}  has found that the process of creatively fusing two things is cognitively demanding.
In formative interviews with designers, researchers found ideation exhausting, especially when listing diverse associations needed to find a match. 
Amateurs found this even more difficult because they lack training to brainstorm diverse associations. 
When evaluating their systems, researchers found that supporting creative tasks with tools ~\cite{metamap, visiblends, symbolfinder} increased the number of design outputs and satisfaction of amateurs, and helped them come up with new ideas that avoided clichés.
However, none of these systems focus on using pop culture as one of the input spaces, which is a special type of conceptual blending because it relies on scenes from a more limited domain.

\enlargethispage{12pt}

In pop culture campaigns such as Star Wars Day, over five thousand brands posted tweets related to Star Wars and their brands in 2018 with hashtags such as \#{starwarsday}~\cite{swday_news,swday_statistics}. 
When we scraped Star Wars posts in 2020, we found over one hundred examples of pop culture blends such as those in Figure \ref{fig:formative_study_samples}.
More than simply having Darth Vader hold a cup of ice-cream, these brands use conceptual blending to integrate their product into memorable pop culture scenes---like Luke using the force to extract his lightsaber (or french fry) from ice (or ice-cream).
Thus, pop culture blends are a creative way to associate two seemingly unrelated things in a meaningful way.

\subsection{Creativity Support Tools for Divergent and Convergent Thinking}
There are many systems that have taken promising approaches to computational creativity and creativity support. 
Many structure the divergent and convergent thinking processes and some use AI or other computational techniques to assist people. 
Two recent systems~\cite{visiblends, metamap} address the related problem of helping people create visual blends. Visual blending is a graphic design technique that draws attention to a message. 
The input is two words that should be connected, and the goal is to create an image that blends symbols of the words together in a way that both symbols are integrated but still individually recognizable. 
VisiBlends~\cite{visiblends} structures this process by expanding both concepts into visual symbols associated with each concept either by brainstorming or with computational tools ~\cite{symbolfinder}, then computationally searching for symbols that can be combined based on having similar shapes. 
MetaMap~\cite{metamap} is an ideation tool for visual blends that leverages the powers of exemplars in design and allows users to search for inspiration based on visual and semantic features extracted from examples. 
It also uses a word association database ~\cite{swow} as a source of associations for divergent thinking and allows users to search over visual and semantic features to find matches.
Blending is also an important aspect of the creativity process in other domains such as creating icons ~\cite{iconate} and fashion design ~\cite{fashionq}.  
These blending tools (as well as the visual blending tools) help in the difficult divergent and convergent thinking process, but still require users to spend time and attention guiding the process for divergence and convergence. 

Divergent thinking is difficult for people because it requires recalling multiple diverse associations from memory.
Many computational approaches have been shown to aid this process.
Knowledge graphs have been shown to support brainstorming by helping people recall associations between concepts~\cite{inspirationwall, spinneret}. 
Word embeddings~\cite{v8storming} such as GloVe~\cite{glove} and Word2Vec~\cite{word2vec} can also be used to support divergent thinking, sometimes in combination with other knowledge graphs~\cite{metaphoria}.
Crowdsourced association data such as Small World of Words (SWOW) ~\cite{swow} has shown to be useful in aiding brainstorming ~\cite{metamap,symbolfinder} and has many advantages over word embeddings such as containing more specific associations~\cite{swow_better_w2v}. 
Presenting images to users during a brainstorming can further spur associations ~\cite{ideaexpander,ideawall} and those images can be optimized for preference ~\cite{mayai} or adapted to cultural contexts~\cite{cultural_image_brainstorming}.
While all these approaches are helpful for supporting divergent thinking, a common limitation is that they do not contain enough pop culture knowledge to suggest diverse associations related to pop culture entities such as characters and plot elements.
Consequently, we explore the use of GPT-3~\cite{gpt}, which has rich knowledge of pop culture through its training data of large-scale Internet text. 
Building upon prior works, we also use SWOW words and images to help users form divergent associations.

Convergent thinking is difficult for people because it requires synthesizing many pieces of information into a cohesive output which has very high cognitive load. 
Computational techniques for convergent thinking almost always involve searching over data to find connections. 
Creative design tasks are ill-defined, and thus there is no exact formula to converge on a solution, but there are several  ways to help ease the cognitive load. 
Design patterns are computational approaches to help uses synthesize elements into a solution~\cite{maneeshdesignprinciples}.
If the design goal can be expressed computationally, design patterns can serve as a constraint-based search to fully automate the search process. 
This has been shown to work for video editing ~\cite{peggy_auto_instr_video,leake_video_editing}, interior design~\cite{interior_design}, making maps~\cite{line_drive}, or assembly instructions~\cite{stepbystepassembly}.
When the constraints are not fully known, the synthesis can be semi-automated with a human in the loop. 
This has been shown to work in visual blends,~\cite{visiblends}, story creation ~\cite{motif}, human-robot interaction~\cite{design_patterns_prototyping_HRI, design_patterns_social_HRI} and finding metaphors~\cite{metaphoria} and analogies~\cite{inventingwithcrowds, analogicalideas}. 
In PopBlends, we present three different methods for searching for connections between the inputs.
Together, these methods form distinct design patterns towards finding associations between two different domains. 
To automate the search, PopBlends formulates the design goal as the semantic similarity of possible connections as represented as sentence embeddings~\cite{sbert}.

Natural language processing tools are powerful tools for creativity support because they allow us to mine free-text created by users (or scraped from the Internet) into structured data that can be further processed. 
Systems like Crosspower~\cite{crosspower} and  Crosscast~\cite{crosscast} parse natural language to generate visual presentations for text. 
They leverage the structure of text that helps search for images and map to visual transitions. 
In parsing free-form text, co-referencing and semantic role labeling are important NLP tools that now are accurate enough for this task. 
In PopBlends, we also parse free-text of plot summaries as a source of entities from pop culture domains to quickly and easily create a domain-specific knowledge base.

\subsection{Large Language Models for Creativity Support}
Large, pretrained language models such as GPT-3~\cite{gpt} denote a fundamental shift in modern NLP. 
Due to their immense size and scale of training data, such models are able to perform well on diverse language tasks without the need for task-specific architectures or fine-tuning. 
Several properties of large language models (LLMs) make them ideal creativity support tools. 
First, LLMs have been found to capture commonsense knowledge that is present in training data, making them effective knowledge bases that are easy to query with natural language prompts. 
For example, results show that BERT~\cite{bert} is competitive with traditional information extraction methods of constructing knowledge bases, as well as open-domain question answering~\cite{gpt_as_kb}. 
Recent work has also explored the potential of using GPT as an open knowledge graph \cite{wang2020language}, motivated by the fact that these language models have been pre-trained to gain knowledge from Internet-scale corpora, which cover the field of pop culture. 
Second, generative language models such as GPT-3 are able to produce machine-generated text that is almost always of high enough quality to seem written by a human. 
These generative capabilities have enabled GPT-3 to aid or complete creative tasks, such as providing ideas for movie scripts, or writing short stories.
Finally, LLMs have often been found to produce hallucinations---which are factually incorrect, but statistically probable, outputs. 
While this is problematic for many domains, such as chatbot assistants or news article generation, recent studies have found that for creative domains---such as story writing---hallucinations can potentially help the user, who is able to reference the output without relying on its correctness~\cite{wordcraft}.

Recent papers have explored many potential uses for LLMs as creativity support tools. 
The writing assistant Sparks~\cite{gero_sparks} showed that GPT helped science writers with three tasks: providing interesting angles to engage readers, crafting concise sentences, and providing reader perspectives. 
The WordCraft~\cite{wordcraft} system explored how a LLM can be used in open-ended ways to support fiction writers in multiple tasks such as ``rewrite this text to be more Dickensian'' or providing suggestions to overcome writer's block. 
Both these writing systems found that writers thought text generation to be useful even when the output was not perfect. 
BunCho~\cite{buncho} is a system that assists writers in generating titles and synopses from keywords. 
AI Chains~\cite{ai_chains} is a tool and technique for combining outputs from a LLM to complete larger tasks like elaboration on feedback to make it more detailed. 
In game generation, GPT-3 has been used to generate interactive stories by restricting how many characters it can introduce (too many characters creates stories that are too hard for a person to follow)~\cite{ai_dungeon}. TaleBrush~\cite{talebrush} allows users to draw a story arc and GPT generates a story that follows that arc. 
Together, these papers show the multiple ways LLMs can support creativity. 
In PopBlends, we focus specifically on supporting divergent and convergent thinking and test three different strategies for combining GPT-3 with traditional NLP technologies.
\section{PopBlends System}

\begin{figure*}
\centering
\includegraphics[width=0.9\textwidth]{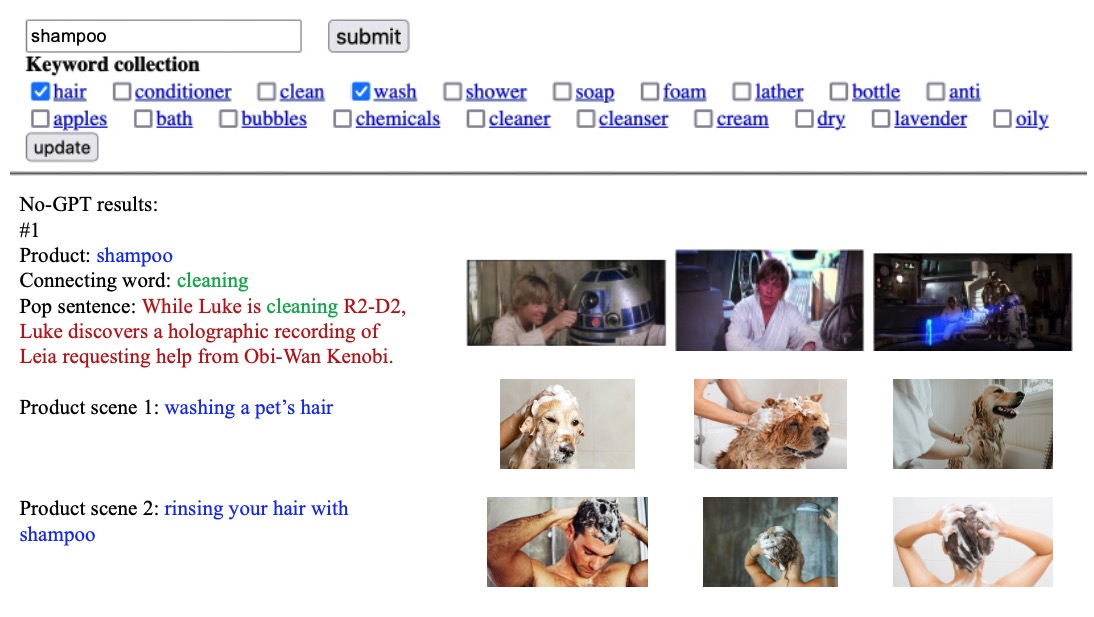}
\caption{PopBlends system screenshot of No-GPT results with Star Wars and shampoo plus selected related words ``hair'' and ``wash'' (Star Wars-related images \textcopyright Lucasfilm Ltd.).}
\label{fig:system_snapshot}
\end{figure*}

The PopBlends system (see Figure \ref{fig:system_snapshot}) is an automated two-stage pipeline to suggest a conceptual blend for two inputs: a pop culture domain and a product or service the user wants to promote.
In Stage 1, the automatic algorithm uses three approaches for finding connections between the inputs: No-GPT, Half-GPT and Full-GPT. 
In each approach there is an initial divergent step that expands domains into associations, and an initial convergent step to find connecting concepts that unite the two domains. 
Once the connecting concept has been found, a second divergent and convergent step is performed in Stage 2 to find scenes and images from both domains related to the connecting concept for blending.

The target users of PopBlends system are amateur designers.
While professionals make many of the pop culture blends for brands seen on social media, amateurs also have things they want to promote---school clubs, nonprofits and community organizations that cannot afford professionals. 
Design tools are especially for amateurs who lack the training and experience in the design process involving divergent and convergent thinking.

\subsection{System Inputs}
There are two inputs in PopBlends: (1) a pop culture domain, and (2) a product or service the user wants to promote (which we will simply refer to as the product). 
Our implementation supports five pop culture domains that users can select from, which are all films or TV shows.
However, we can easily add more with an automated pipeline. 

For the product domain input, users can give the system any product words they want. 
However, one problem is that many product words have multiple meanings.
For example, a cookie is both a food and a file related to browser data. 
To ensure that the system correctly understands the user's intent, we display a selectable list of related words alongside the original word. 
These related words are automatically fetched through querying the Small World of Words (SWOW) \cite{swow_better_w2v} knowledge graph. 
The user then selects the words that can clearly identify the product, thereby disambiguating their intent. 
Finally, the system combines the semantic information contained in the original product word and user-selected related words in a single semantic representation which we call the \textit{product embedding}.

\subsection{Stage 1: Find Connecting Concepts}
\begin{figure*}
\centering
\includegraphics[width=1.0\textwidth]{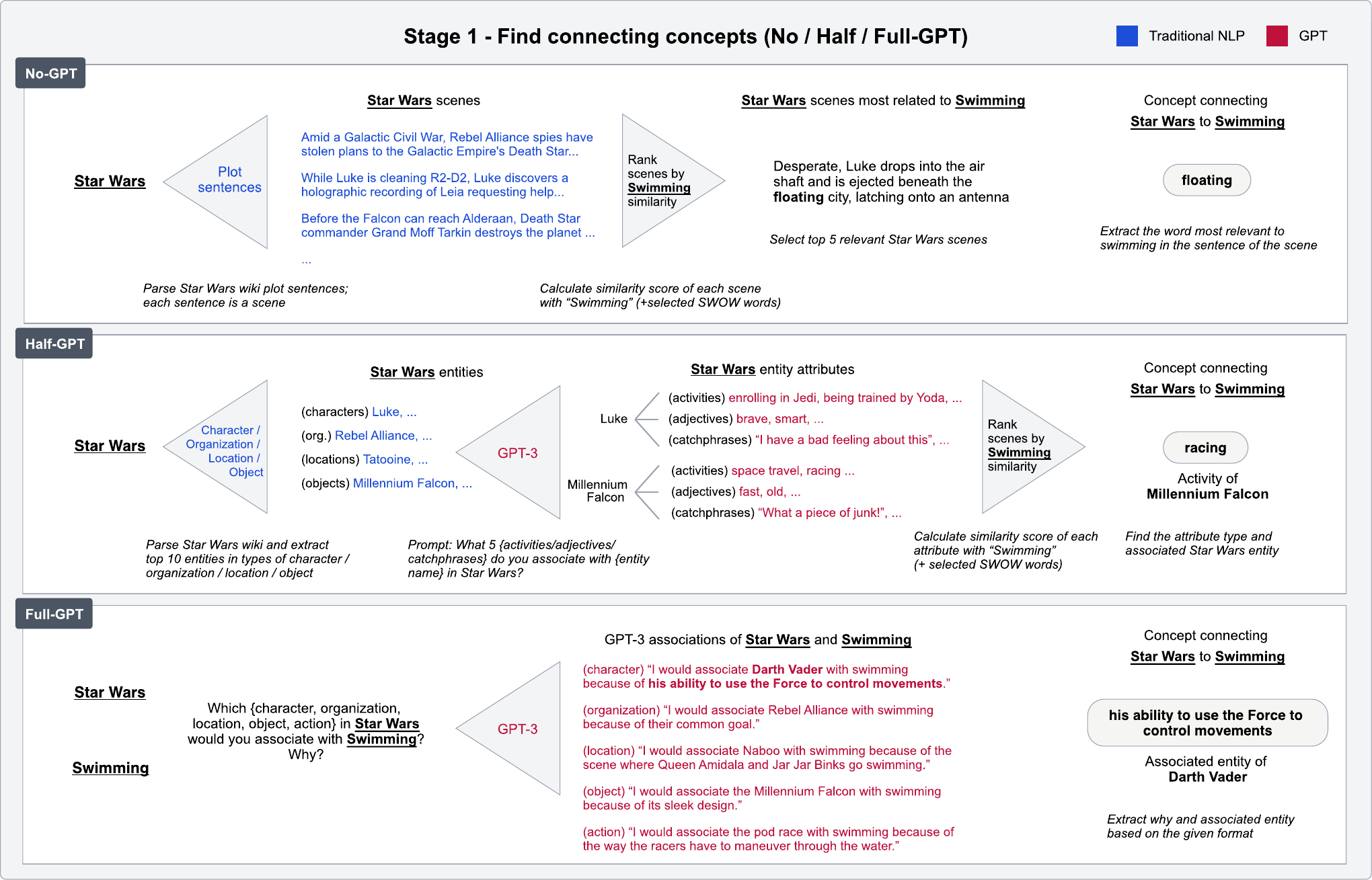}
\caption{PopBlends Stage 1 system diagram with inputs of Star Wars and swimming, where we explore three different strategies to find connecting concepts: No-GPT, Half-GPT, and Full-GPT.}
\label{fig:stage1_diagram}
\end{figure*}

In Stage 1, the system's goal is to find \textit{connecting concepts}, which are intermediate ideas that link the pop culture and product domains. 
As shown in Figure \ref{fig:stage1_diagram}, we explore three distinct methods to find connecting concepts: No-GPT (traditional NLP and knowledge extraction only), Half-GPT (combining knowledge extraction and GPT), and Full-GPT (relying solely on GPT for connections).  
Each method performs a divergent step to expand each domain into a large list of candidate concepts; next, a convergent step is performed to select the final connecting concepts from the candidates.
In effect, each method produces potentially distinct connecting concepts, which are jointly presented to the user through the web interface. 
More examples of connecting concepts across pop culture domains can be seen in Table \ref{tab:airplane_examples}.

\subsubsection{No-GPT}
The No-GPT approach uses traditional NLP and knowledge extraction to find connecting concepts. 
Specifically, we rank scenes from the plot of the pop culture domain (e.g., Star Wars) based on their semantic similarity to the product domain (e.g., Cookies), then extract the single most relevant word from highly ranked scenes as connecting concepts. 
We focus on pop culture scenes due to their prominence in pop culture blends---as Figure \ref{fig:formative_study_samples} shows, pop culture blends typically depict important events that occur between characters in a show. 
For example, to create the blend of Star Wars and Girl Scout cookies, we need to know the scene that involves Luke (the light side of force) fighting Vader (the dark side of force). 

To create a database of pop culture scenes, we scrape the ``Plot'' section of the Wikipedia page(s) for each domain.
This provides us with information on the most important characters, organizations, locations, actions and events in each movie or TV show. 
Alternative approaches were considered, such as scraping data from fan-compiled data sources like Fandom Wiki pages. However, we found that they cover too many minor details (minor characters, obscure back stories) and mix character trivia with plot summary. 
We also considered Google Knowledge Graph~\cite{google_knowledge_graph}, which can list the main characters, but not any additional associations like their jobs or famous scenes. 
With the data of wiki plot summaries, we then perform the divergent and convergent steps to find connections.

For the divergent step, we expand the scraped pop culture data into a collection of plot sentences, since each sentence typically contains a useful action or scene in which the character is involved. 
We do this by splitting the summary into individual sentences with the NLTK tokenizer ~\cite{nltk}. 
However, one complication that arises is that when plots are parsed into independent sentences, there are many cases where pronouns instead of names are used to refer to entities. 
We use co-reference resolution from the AllenNLP API ~\cite{allennlp} to resolve these co-references and replace all pronouns with their referent.

For the convergent step, we find the most relevant pop culture plot sentences to the product---i.e., the sentences with the highest semantic match to the product embedding.
We use semantic similarity matching techniques based on asymmetric sentence transformers \cite{wang2020minilm} for this task. 
We found that sentence transformer-based similarity outperformed word-based matching, which is comparatively coarse. 
We choose the 5 sentences with the highest matching scores to find a connecting concept from each. 
To do this, we calculate the semantic similarity between the product embedding and each word in the highest matching plot sentence, and return the most-related word. 
For example, in the case of Star Wars and swimming, one of the highest matching plot sentences is ``Desperate, Luke drops into the air shaft and is ejected beneath the floating city, latching onto an antenna'' and the connecting concept is ``floating''. 
Such plot sentence matching is very faithful to the pop culture domain, but misses many potential connections. 
There are lots of types of information plot summaries do not cover, including many entity attributes such as characters' catchphrases. 
Thus, there is a need to explore other approaches beyond using the plot summary as the sole knowledge base.
 
\subsubsection{Half-GPT}
The Half-GPT approach combines knowledge extraction with GPT-3 to find connections based on entity attributes.
In the Half-GPT approach, we first collect a list of important pop culture entities, and then take the list to perform divergent and convergent steps to find connections.
First, we collect a list of important entities---characters, organizations, locations and objects, based on the scraped Wikipedia plot summaries. 
To extract entities from the plot text, we use knowledge extraction technologies---we run an ELMo-based named entity recognition model \cite{peters2017semi} to get the named entities with tags of people, organizations, locations and miscellaneous, which are a large part of entities we want. 
However, we also want non-named entities such as lightsabers in Star Wars. 
To get important non-named entities, we run TF-IDF \cite{ramos2003using} against 100 random Wikipedia pages to find words and phrases unique to the pop culture domain. 
Together, these represent the most important characters, organizations, locations and objects in the domain.
Some pop culture domains like Game of Thrones have long plot summaries that contain a myriad of entities, many of which are not iconic or easily recognizable by the average fan.
Therefore we rank the entities based on their frequency according to TF-IDF and use only the top ten in each category.

For the divergent step, we use GPT-3 to expand the pop culture entities to attribute associations, including activities (Monica and cooking in Friends), adjectives (Chewbacca is hairy in Star Wars) and catchphrases (Joey and ``Joey doesn't share food!'' in Friends).
This information is not present in the plot summaries, and also sparsely available or difficult to systematically scrape from Fandom wikis or other knowledge bases. 
Thus, we treat GPT-3 as a knowledge base and query it to get these attributes. 
Based on the ranked entity list, we use the GPT-3 API to ask \textit{``What five \{activities, adjectives, catchphrases\} do you associate with \{entity name\} in \{pop culture domain\}?''}. 
We parse the free-text responses from GPT-3 into attributes and cache them.
Overall, this results in 600 attributes (4 entity types $\times$ top 10 entities per type $\times$ 3 attribute types $\times$ 5 attributes per type) for each pop culture domain.

For the convergent step, we use the semantic similarity matching techniques \cite{wang2020minilm} to find the 5 entity attributes that best match the product embedding. 
For each matching attribute, we find its associated entity. 
For example, for Star Wars and swimming, a matching attribute is ``racing'', and it is associated with the entity Millennium Falcon.
If there are multiple entities that have the same attribute, we only return the top two most popular associated entities for that concept. 
For example,  in Game of Thrones, ``fighting'' is associated with almost all the characters. 
But we want to return a diverse set of connecting concepts, so we only return ``fighting'' (and its two most associated entities: Jon Snow and Daenerys Targaryen) once. 

\subsubsection{Full-GPT}
In the Full-GPT approach, we investigate whether GPT-3 can make the connection between the product and the pop culture domain on its own.
For the divergent step, we prompt GPT-3 with 5 different questions. 
We ask GPT-3 which character, organization, location, object and action it can associate with the product term. 
We ask GPT-3 to answer in a format so that we can extract easily. 
Here is the exact prompt we use: \textit{``Which \{character, organization, location, object, action\} in \{pop culture domain\} would you associate with \{product term\}? Why? Please say your answer in the format of ``I would associate ... with ... because of ...''}. 
Answers to this prompt usually contain both a matching entity and a justification. 
For example, ``I associate swimming with Darth Vader because of his ability to use the Force to control the movements.'' 
GPT-3's answers may be factual-based or not. 
For example, for the combination of Star Wars and boxing, GPT-3 answers ``I associate boxing with Jabba the Hutt's palace because of its underground fighting ring.'' 
To our knowledge, Jabba the Hutt does not have an underground fighting ring.
However, this answer could still be useful. 
One could imagine a blend of putting two boxers into the fighting ring of Jabba the Hutt's palace for a match.

\subsection{Stage 2: Find Scenes Based on the Connecting Concept}
\label{system_stage2}
\begin{figure*}
\centering
\includegraphics[width=0.88\textwidth]{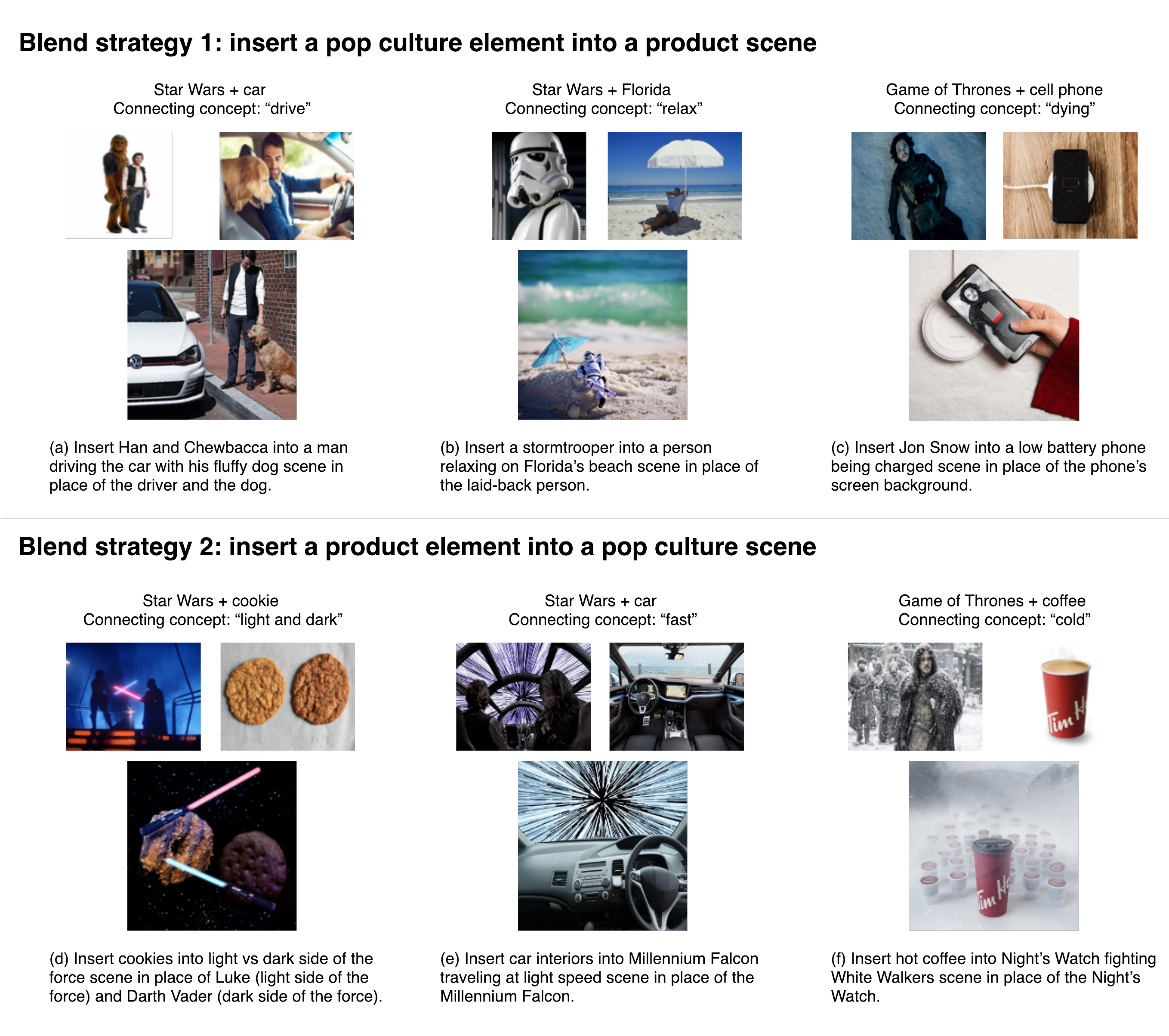}
\caption{Professional blend examples of Star Wars and Game of Thrones, collected on Twitter from a. Volkswagen, b. Visit Florida, c. Samsung, d. Girls Scout, e. NSW Police Force, f. Tim Hortons, where a--c follows the blend strategy to insert a pop culture element into a product scene, d--f inserts a product element into a pop culture scene (Star Wars-related images \textcopyright Lucasfilm Ltd., Game of Thrones-related images \textcopyright Home Box Office, Inc.).}
\label{fig:prof_images}
\end{figure*}

At the end of Stage 1, we get connecting concepts across the two domains, such as ``racing'', which connects Star Wars and swimming. 
However, only knowing the connecting concept is not enough to make a blend.
In Stage 2, we need to find the context and images to implement the connecting concept generated in Stage 1 into a blend. 
By analyzing the professional blends (See Figure \ref{fig:prof_images}), we find two ways of making pop culture blends for a connecting concept: 1) insert a pop culture element into a product scene (See examples a--c) , 2) insert a product element into a pop culture scene (See examples d--f).
Typically, the inserted element replaces something similar or analogous in the scene, for example, in example c, the dying Jon Snow is inserted to replace the screen background of a nearly dead cell phone.

\begin{figure*}
\centering
\includegraphics[width=1\textwidth]{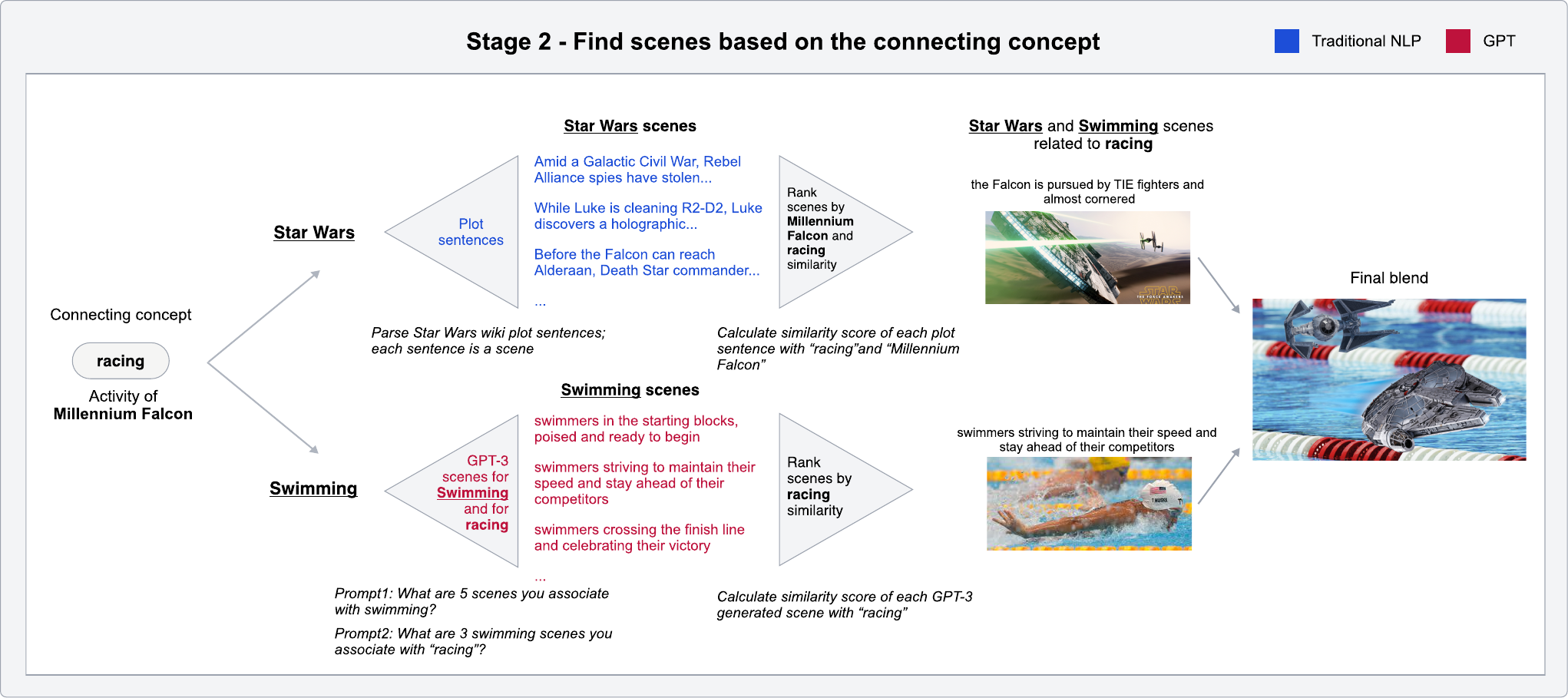}
\caption{PopBlends Stage 2 system diagram with the connecting concept of ``racing'' between Star Wars (\textcopyright Lucasfilm Ltd.) and swimming, where we first expand both pop culture and product domains into scenes and then find the scenes and their images that are relevant to the connecting concept (Star Wars-related images \textcopyright Lucasfilm Ltd.).}
\label{fig:stage2_diagram}
\end{figure*}

To do the insertion, we need related scenes and images centered on the connecting concept in either pop culture or product domains. 
Therefore, in Stage 2 (see Figure \ref{fig:stage2_diagram}), we first take the divergent step to expand both pop culture and product domains into scenes and then take the convergent step to find the scenes and their images that are relevant to the connecting concept.
For example, based on the concept ``racing'' (activity of millennium falcon) that connects Star Wars and swimming, to make a blend, we need to find Star Wars scenes related to ``racing'' (and millennium falcon) as well as swimming scenes related to ``racing''.

\subsubsection{Find pop culture scenes related to the connecting concept}
For the divergent step, we expand the pop culture domain into Wikipedia plot sentences, which include rich pop culture scenes. 
For the convergent step, we first find relevant pop culture scenes---for the Half-GPT and Full-GPT connecting concepts, we find the two most relevant plot sentences to the connecting concept and its associated entity with semantic similarity matching techniques \cite{wang2020minilm}; for the No-GPT connecting concepts, we simply take the most relevant plot sentence (already found in Stage 1). 
Then, we find three images with each plot sentence as a query term in the Google Image API.

\subsubsection{Find product scenes related to the connecting concept}
For the divergent step, we use GPT-3 to expand the product term into associated activities and their contexts. 
This data is collected by querying GPT-3 with prompts in the form: \textit{``What are five scenes you associate with \{product term\} ?''}. 
For example, for swimming, GPT-3’s response is ``1) swimmers diving into a pool 2) swimmers doing laps in a pool 3) swimmers competing in a swimming race 4) swimmers playing in the water 5) swimmers enjoying the water on a hot day''. 
In addition to collecting the general product-related scenes, we also expand product domain into connecting concept-related scenes by asking GPT-3 with the prompt of \textit{``What three \{product term\} scenes do you associate with \{connecting concept\}?''}.
For example, for ``racing'' related swimming scenes, GPT-3’s response is ``1) swimmers in the starting blocks, poised and ready to begin 2) swimmers striving to maintain their speed and stay ahead of their competitors 3) swimmers crossing the finish line and celebrating their victory''. 
We split these responses into individual sentences and store them for the next phase. 

For the convergent step, we find the two product scenes that are most related to the product embedding and the connecting concept using semantic similarity matching techniques \cite{wang2020minilm}, and find three images for each of these scenes using Google image API.
\section{Technical evaluation}
We conducted an annotation study to evaluate the quality of PopBlends system outputs in Stage 1: connecting concepts and Stage 2: scenes based on the connecting concept. 
For Stage 1, we rate how often the provided connecting concepts are relevant to both the pop culture and product domains. 
We compare the results across the three GPT strategies to answer one of our research questions---which GPT strategy is the best. 
For Stage 2, we evaluate if the expanded pop culture or product scenes are relevant to the connecting concept. 

\subsection{Pop Culture and Product Topic Collection}
\label{topic_selection}
We provide annotators results for combinations of 5 pop culture domains and 6 products. 
The 5 pop culture domains are Star Wars, Friends, Harry Potter, Game of Thrones and Breaking Bad, which are all well-known and used to create pop culture blends on social media. 
These domains also cover various genres and settings of pop culture, including movies and TV shows with fictional and real-world settings.
The products were randomly picked from the six most common categories in a visual advertisements dataset \cite{ad_symbols}: 1) clothing, accessories, beauty products and cosmetics; 2) cars, automobiles; 3) food (restaurants, chocolate, chips); 4) drinks (soda, alcohol, coffee, tea); 5) electronics; 6) sports equipment and activities. 
The 6 randomly selected products were: 1) \textit{toothbrush}; 2) \textit{airplane}; 3) \textit{pizza}; 4) \textit{beer}; 5) \textit{cell phone}; 6) \textit{swimming}.

\subsection{Evaluating Stage 1: Are the connecting concepts related to both the pop culture and product domains? }
\subsubsection{Methodology}
\label{annotators}
We recruited 10 annotators (7 female; mean age 23.5, std 2.29), two for each pop culture domain, through a university mailing list and Facebook group. 
Each participant was required to have familiarity with their pop culture domain.
For the 15 connecting concepts (5 for each strategy---No-GPT, Half-GPT, Full-GPT) for each of the pop culture-product pairs, we ask the annotator two questions: Q1) ``Is the connecting concept related to the pop culture domain?'' and Q2) ``Is the connecting concept related to the product domain?'' 
For both questions, two annotators independently annotate true or false. The inter-rater reliability shows that there is moderate agreement between the two annotators’ judgments with Cohen’s $\kappa = .43$.  
It takes a bit of imagination to determine whether a connecting concept is related to a product or pop culture domain, and our raters were conservative in that some did not see the connection while others did.
For example, in determining whether the connecting concept of ``flying in a spaceship'' is related to airplane, one annotator saw the connection while the other did not make the cognitive leap between the spaceship and airplanes.
Hence, if at least one of them annotates true, we mark the answer to this question as true, otherwise it is false. 
We then mark the success of each connecting concept by whether the answers to both questions are true.

\subsubsection{Accuracy of Stage 1 across the three GPT strategies}

\begin{figure*}
\centering
\includegraphics[width=0.6\textwidth]{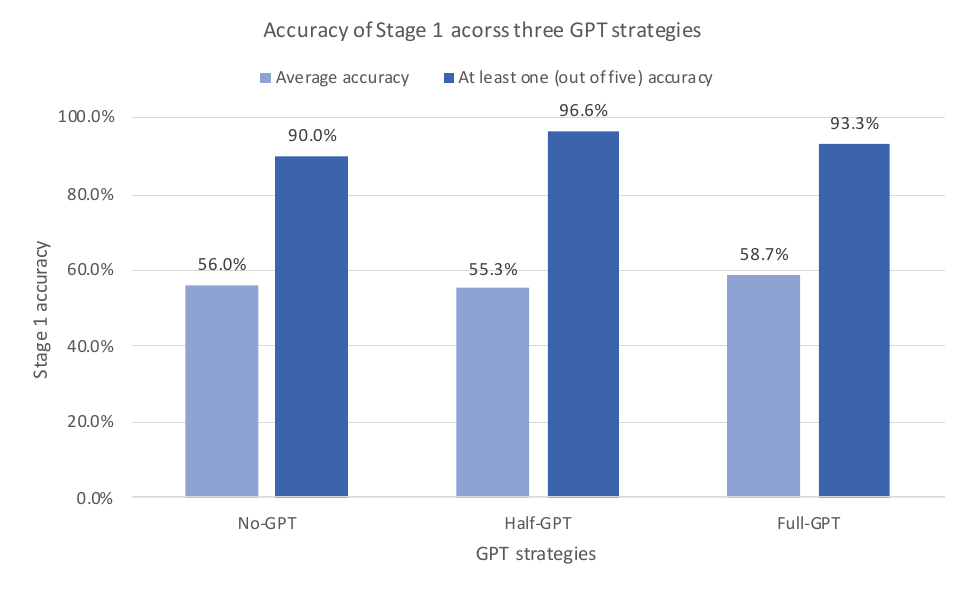}
\caption{Average accuracy and at least one out of five accuracy of Stage 1 across three GPT strategies}
\label{fig:stage1_accuracy}
\end{figure*}

Overall, across all three GPT strategies, 56.7\% of connecting concepts were related to both the pop culture domain and the product domain. 
Moreover, 93.3\% of all pop culture-product pairs had at least one good connecting concept.
These are encouraging results because in previous work ~\cite{opal, symbolfinder}, researchers demonstrated the high cognitive load of memory recall tasks, with 6 out of 7 on the NASA-TLX dimension of mental demand ~\cite{opal}. 
Because recognition is so much easier than recall, users are willing to tolerate some machine errors and they still find the system useful because it lowers cognitive load.
Having a 93\% chance of at least one good result demonstrates consistent value of automated tools.

Surprisingly, all three GPT strategies performed equally well. 
The success rates of No-GPT, Half-GPT, and Full-GPT strategies were 56.0\%, 55.3\%, and 58.7\%, respectively. 
For each GPT strategy, we annotated 5 different connecting concepts; the probability of success for at least one of the five connecting concepts was 90.0\%, 96.6\%, and 93.3\%, respectively (See Figure \ref{fig:stage1_accuracy}).
This shows all three GPT strategies are capable of producing at least one good connecting concept in 5 chances. However, we find that the GPT strategies produced surprisingly different results. 

The No-GPT strategy is accurate when the plot sentences contain a concept directly or indirectly related to the product. 
For example, for products \textit{airplane} and \textit{cell phone}, ``airplane'' and ``phone'' are directly used in Friends plot sentences and the No-GPT strategy was able to find them as the connecting concepts. 
Also, the No-GPT strategy can find indirect associations with products like ``floating'' (Star Wars + \textit{swimming}) and ``airport'' (Friends + \textit{airplane}). Occasionally, the No-GPT approach can find connecting concepts that are distantly related to the product but still useful, such as ``using''  (Star Wars + \textit{toothbrush}), in which case we can make a blend by replacing the lightsaber with a toothbrush in the scene of using the force.
However, sometimes the No-GPT approach does not work well because the connecting concept could be matched on incorrect word sense across two domains, for instance, ``solo'' connects Star Wars (Han Solo, a main character in the movies) and \textit{swimming} (solo swimmer), ``crew'' connects Breaking Bad (gang crew) and \textit{airplane} (airplane crew). 
Note there is a potential of making a blend here, but it is based on the pun rather than the visuals. 
For example, an airline ad could have a slogan like Breaking Bad ``crew'' serving as the flight attendants. 
In addition, as the No-GPT strategy prioritizes the pop culture domain, the connection of the connecting concept to the product may be weak, such as ``money'' (Breaking Bad + \textit{pizza}).

The Half-GPT strategy is based on entity attributes rather than plot sentences and produces different results from the No-GPT strategy. 
The Half-GPT strategy works even if there is no plot sentence that directly describes the pop culture entities’ attributes.  
For example, 1) as an activity of Millennium Falcon, ``racing'' connects Star Wars and \textit{swimming}, and a possible blend could be Millennium Falcon racing in a swimming pool like a speedboat, with TIE fighter behind it; 2) as an adjective of Tohajiilee Indian reservation, ``remote'' connects Breaking Bad and \textit{cell phone}, where a possible blend could be using the brand’s cell phone from the remotest of location with the slogan ``Finally, a phone that works where you need it to.''
And advertisers can show a picture of Walter White using the phone in the reservation in his underwear; 3) as a catchphrase of Joey, ``Joey doesn't share food'' connects Friends and \textit{pizza}, and a possible blend might be Joey concentrating on eating pizza from the brand that needs to be advertised, with the tagline ``Joey doesn't share this brand's pizza!''
As GPT-3 sometimes will provide weak attribute connections, Half-GPT may fail, especially when the entities are not significantly important in the pop culture domain. 
Note that we measure the overall accuracy of GPT-3 generated entity attributes in a separate annotation study in Section \ref{attribute_accuracy_study}. 
For example, as a trait of Vermont, ``perfect'' connects Friends and \textit{airplane}, where the similarity score between \textit{airplane} and ``perfect'' is 0.16; as a trait of Whomping Willow, ``big'' connects Harry Potter and \textit{beer}, where the similarity score between \textit{beer} and ``big'' is 0.17. Note that we forced PopBlends to show the top five connecting concepts for each strategy without a similarity score cut-off. 
To get rid of these bad cases, we can set up a cut-off score or detect if the similarity score has a dramatic drop compared to that of the higher ranked connecting concepts.

Note that the Half-GPT strategy relies on traditional NLP techniques to collect entities from plot summaries.
These traditional NLP techniques, such as co-reference resolution~\cite{Lee2018HigherOrderCR} and entity extraction~\cite{peters2017semi}, have already been shown to be highly accurate. 
We corroborate that they are also highly accurate in the context of pop culture. 
Table \ref{tab:pop_entities} shows a collection of the important entities for each of the five pop culture domains.
We compared these entities with those on  Fandom wikis, and we verified all (100\%) of the collected entities were clearly relevant to the pop culture domains with no repetition. 
4\% (8 out of 200) entities are potentially in the wrong category, for example, Admiral Ackbar should be a Star Wars character rather than organization.
This is probably because NLP techniques we used does well in entity recognition but a little less in role tagging. 

In the Full-GPT strategy, we ask GPT-3 to come up with the connections on its own. 
We find that, although sometimes the results have some overlap with those of the No-GPT and Half-GPT methods, in general the Full-GPT approach can provide different connecting concepts that often go beyond the limits of the collection of plot sentences and entity attributes (adjectives, activities, catchphrases). 
The connecting concepts provided are sometimes surprising and about physical appearance. 
For example, \textit{toothbrush} has ``the long and thin handle'', as does a lightsaber in Star Wars; ``Quidditch cup’s use as a prize'' connects Harry Potter and \textit{beer} because Quidditch cup and beer are both cup-sized and golden in color ; C-3PO's ``shiny exterior'' connects Star Wars to \textit{shampoo} as it describes the texture of the hair after being washed by the shampoo.
However, full use of GPT-3 also brings disadvantages. 
First, GPT-3 may fail to make a connection to the pop culture domain, although it claims to. 
For example, GPT-3 claims ``the connection between using a toothbrush and cleaning one’s teeth'' as a connecting concept between Star Wars and \textit{toothbrush} and says that ``the connection between pizza and being content'' could connect through Star Wars to \textit{pizza}. 
Second, GPT-3 is prone to hallucination, which in our case means claiming something that never exists in the pop culture domain. 
For instance, GPT-3 generated connecting concepts of ``meth phone and cell phone’s shared portability and lack of wire connections'' for Breaking Bad and \textit{cell phone}, and ``Anakin Skywalker's death in water'' for Star Wars and \textit{swimming}. 
Occasionally, these hallucinations can inspire blends, but it is rare. 
For example, GPT-3 lists ``Phoebe's smelly cat fundraiser's constant use of social media to spread awareness'' as one of the connecting concepts between Friends and \textit{cell phone}, which is imaginable and connects cell phone well to Phoebe’s smelly cat song.

\subsubsection{Accuracy of entity attributes in Half-GPT}
\label{attribute_accuracy_study}
The Half-GPT strategy uses GPT-3 as a pop culture database to collect three different types of entity attributes---activities (``fights for the Galactic Empire''), adjectives (``agressive''; ``evil'') and catchphrases (``I am your father''), with example attributes from Darth Vader in Star Wars. 
To evaluate GPT-3's performance in this task, we conduct a separate annotation study to compare GPT-3's accuracy in producing these three types of attributes for a variety of entities in each pop culture domain.

We recruited 10 annotators (8 female; mean age 23.7, std 2.16) who were experts in each of the pop culture domains. 
We use the same recruiting channels as the previous annotation study; six of the annotators participated in both studies. 
For each of the collected attributes, we asked two annotators to independently answer ``How many of the items (out of 5 GPT-3 answers) are relevant to the given entity?'' with an integer in the range of 0 to 5. 
Note that the interrater agreement (0.41 as measured by Pearson’s $r$) is relatively low between pairs of annotators. 
This highlights the subjectivity of ``relevance'' between attributes and domains—two concepts may feel relevant to some people but not to others.

Overall, the average of the attribute accuracy is 2.86 out of 5 (57\%). 
The most accurate attribute type provided by GPT-3 was adjectives  (79\%).
Activities were less accurate (58\%), and catchphrases were the least accurate (35\%)
(See Table \ref{tab:attribute_relevance}). 
GPT-3 was able to produce correct adjectives for all entity types---characters (Darth Vader is aggressive), organizations (Rebel Alliance is heroic), locations (Mos Eisley is dirty) and objects (Imperial fleet is massive).
However, whereas many adjectives were accurate, they were not always the most salient attribute for the entities.
For example, one of Chewbacca's adjectives in Star Wars is ``loyal'', and although this is true, ``loyal'' is not his most prominent characteristic.
Catchphrases have the lowest accuracy, as they are often made up by GPT-3. 
This is in a large part due to the fact that many entities in a movie or TV show do not have classic catchphrases, such as Dementors in Harry Potter, who do not speak. 
Overall, the study demonstrates the effectiveness of GPT-3 as an associative pop culture knowledge base, and the corresponding potential of PopBlends to expand a pop culture domain with associated attributes. 

\begin{table}[]
\begin{tabular}{llrr}
\toprule
\textbf{Attribute} & \textbf{Example output(s)} & \textbf{Avg. (\%)} & \textbf{IRR} \\
\midrule
\textit{Activities} & fights for the Galactic Empire & 2.89 (58\%)   & 0.32  \\
\textit{Adjectives} & aggressive; evil & 3.93 (79\%)   & 0.47  \\
\textit{Catchphrases} & ``I am your father'' & 1.75 (35\%)   & 0.41  \\ 
\midrule
\textit{All} &       & 2.86 (57\%)   & 0.41  \\ 
\bottomrule
\end{tabular}
\caption{Annotated relevance results and example outputs for Darth Vader in Star Wars. \textit{Avg.} is the average count (out of 5) of relevant attributes across all pop culture domains. IRR denotes the correlation (Pearson's $r$) between annotator pairs.}
\label{tab:attribute_relevance}
\end{table}

\subsubsection{Conclusion}
Overall, the probability of success for at least one of the five connecting concepts was at or above 90\% in all three methods. 
The three methods are equally good and each has unique strengths and weaknesses: No-GPT is accurate, but may sometimes match on incorrect word senses; Half-GPT works well even without directly related plot sentences, but GPT-3 sometimes provides weak attribute connections; Full-GPT can provide surprisingly good results, but GPT-3 may pretend to make connections or be prone to hallucinations. 
Therefore, it makes sense to present them all to users, rather than just presenting one type of method. 
We discuss the potential of using the ensemble method in Section \ref{ensemble_learning}.

\subsection{Evaluating Stage 2: Are the pop culture or product scenes related to the connecting concept?}
\subsubsection{Methodology}
For 15 PopBlends outputs (each of them contains a connecting concept, 1 or 2 pop culture scenes, 2 product scenes) of each pop culture-product pairs, we ask the annotators (same people from Section \ref{annotators}): Q3) ``Is the pop culture scene related to the connecting concept?'' and Q4) ``Is the product scene related to the connecting concept?'' 
For each question, two annotators independently annotate either true or false. 
The inter-rater reliability shows that there is moderate agreement between the two annotators’ judgments with Cohen’s $\kappa = .48$. 
Similar to evaluating Stage 1, annotators need to make a cognitive leap to connect the scenes with the connecting concept.
For example, when judging whether the connecting concept ``friendly'' (trait of Ewoks) is related to the Star Wars scene of ``The team ...gaining a tribe of Ewoks' trust after an initial conflict.'' 
While one annotator saw the relatedness, the other did not make the cognitive leap between gaining trust and being friendly. 
Hence, if at least one of them annotates true, we mark the answer to this question as true, otherwise it is false. 

\subsubsection{Difference in relatedness between pop culture and product scenes with the connecting concept}
\label{diff_gen_ways}

\begin{figure*}
\centering
\includegraphics[width=0.6\textwidth]{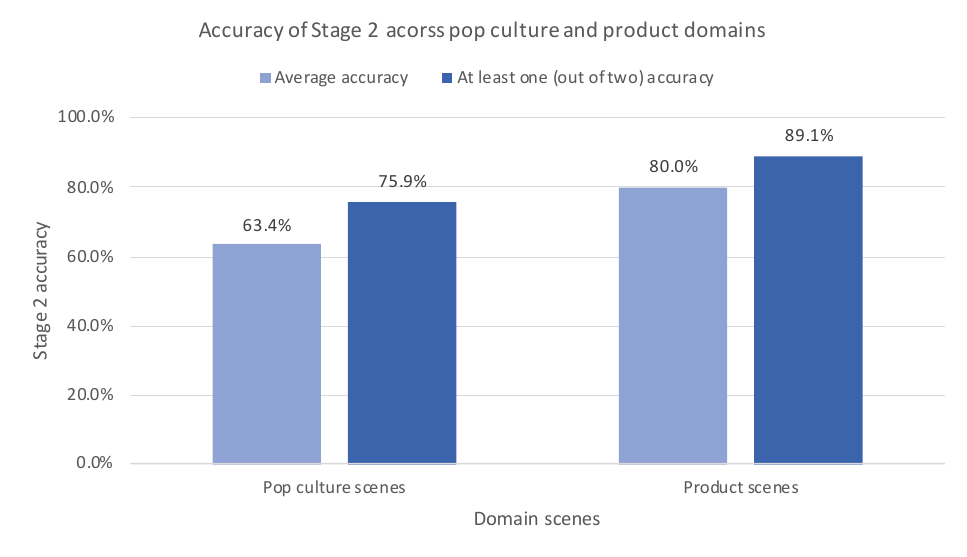}
\caption{Average accuracy and at least one out of two accuracy of Stage 2 for pop culture and product scenes}
\label{fig:stage2_accuracy}
\end{figure*}

Overall, given that the connecting concept from Stage 1 is strong, the probability that at least one scene is related to the connecting concept is 98.3\%. 
This could mean that only the pop culture scene is related, or only the product scene is related or both---but to make a successful blend, we only need one of the two domains to have a related scene (because we can insert an entity from one domain into the scene from another domain, as shown in Section \ref{system_stage2}). 
In general, the product scenes are more likely to be related to connecting concepts than the pop culture scenes.
The probability that at least one product scene is related to the connecting concept is 89.1\%. 
While the probability that at least one pop culture scene is related to the connecting concept is 75.9\% (13.2 percentage points lower). 
This is similar for the raw percentages---the raw relatedness of product scenes is 80.0\% and the raw relatedness of pop culture scenes is 63.4\% (a difference of 16.6 percentage points, see Figure \ref{fig:stage2_accuracy}). 
Also note that for bad connecting words, Stage 2 still works well---there is an average of 92.3\% probability that at least one scene is related to every connecting concept. 
Although we do not expect these scenes to be useful (because the connecting concept is bad), it shows this stage of the pipeline performs almost as well, even with a weak connecting concept. 

The difference of relatedness of scenes from two domains is probably because pop culture and product scenes are generated very differently: the pop culture scenes are retrieved from the database of plot sentences from Wikipedia, whereas the product scenes are generated from GPT-3.
Pop culture scenes are retrieved based on the semantic similarity of the plot sentence to the connecting concept.
Thus, the scenes will definitely be relevant to the pop culture domain, but may or may not be related to the connecting concept. 
If there is no scene in the data of plot sentences with a similar word, then this approach will fail. 
For example, Chewbacca is hairy, but there are no plot sentences directly about that. 
However, ``hairy'' is still a very good connecting concept between Star Wars and \textit{shampoo}, and there are plenty of product scenes that we can insert Chewbacca into to make a final blend, like a before and after picture of shampoo usage.

Meanwhile, product scenes are generated by giving GPT-3 prompts about products and connecting concepts. 
We chose GPT-3 because it is a language model that contains a large-scale corpus of everyday language covering a variety of scenes, which greatly fits the task of generating everyday product-related scenes. 
GPT-3 generally works well. 
For example, when prompted to generate 5 \textit{beer}-related scenes, GPT-3 responded with: ``a guy walks into a bar and orders a beer'', ``a group of friends are sitting around a table drinking beer and chatting'',  ``a couple is sharing a beer while watching a sunset'', ``a guy is drinking a beer while watching a football game on TV'', ``a group of people are having a beer tasting party", which are all great real-world beer scenes to use.
However, GPT-3's outputs are not always great. 
Sometimes GPT-3 may provide completions that consist only of locations or objects, without any people or actions within them. 
For example, for swimming scenes (Star Wars + \textit{swimming}, connecting concept: ``body''), GPT-3’s responses were ``the pool'' and ``the sea''; for pizza scenes (Harry Potter + \textit{pizza}, connecting concept: ``the many different toppings that are available'', associated with the action of choosing pizza as a dinner option in the Hogwarts great hall), GPT-3 generated ``sausage'' and ``green peppers''. 
Next, GPT-3 sometimes provided product scenes that come from other pop culture domains rather than the real world. 
For example, for toothbrush scenes (Star Wars + \textit{toothbrush}), GPT-3 gave an answer of ``the toilet brushing scene from the simpsons''; for pizza scenes (Harry Potter + \textit{pizza)}, GPT-3 answered ``the pizza scene in Friends''.
To improve the GPT-3 product scene generation part, we may need further efforts in prompt engineering. 
For example, we can ask specifically for typical real-world product scenes to avoid getting answers that involve other pop culture domains. 
We may also ask GPT-3 with the same prompt multiple times to get a larger product scene pool, and then we could use NLP techniques to detect if there is a person and an action in the responses to judge their ``concreteness''. 
Based on that, we can add a filtering step for quality control.

Since GPT-3 was able to reliably generate product scenes, we also tried to generate pop culture scenes. 
However, GPT-3 often hallucinated. 
For example, when prompted to generate Star Wars scenes related to ``eaten'' (Star Wars + \textit{pizza}), GPT-3 responded with fabricated scenes like ``Darth Vader eats a human heart'' and ``Leia is pretending to eat a plate of food in order to make Han feel better about himself''. 
To resonate with an audience, a pop culture blend should use real, recognizable scenes. 
Therefore, we employed the original method of similarity ranking within a collection of wiki plots, even though it was more constrained.

\subsubsection{Conclusion}
Overall, the probability that at least one scene is related to the connecting concept is 98.3\%.  
The product scenes (89.1\%) are more likely to be related to connecting concepts than the pop culture scenes (75.9\%), which was probably caused by the different generation approaches.
Generally, GPT-3 works well for generating product scenes but brings hallucinations for pop culture scenes.
Hence, we stick to the wiki database retrieval method for pop culture scene generation in consideration of wiki plot's relatedness to the pop culture domains.
\section{User Study}
To understand if and how PopBlends is useful for creating pop culture blends, we conducted a within-subjects study, comparing PopBlends to general Internet search, which as shown in prior work, is a powerful means for brainstorming ideas and images for concepts \cite{iconate, symbolfinder}. Internet search is useful for creating pop culture blends because there is a large library of online pop culture resources (including texts, images, and videos) that are easy to use as inspiration sources or materials.
Also, users can feasibly use the search tool to augment their memories if they fail to think of any details.
Specifically, we evaluate if users can (1) ideate more pop culture blends with PopBlends and (2) if using PopBlends reduces the perceived difficulty of ideating blends. 

\subsection{Participants}
We recruited 10 students (6 female; mean age 24.2, std 1.14) via email and word-of-mouth at a local university. 
Participants were amateur designers, interested in creating images for social media to promote clubs and products. 
In order to be eligible, participants were required to be familiar with at least one of the pop culture domains we provided.
They were paid 30 dollars for up to 1.5 hours of their time.

\subsection{Procedure}
During the study, the participants' goal was to create six pop culture blends, three with the baseline, three with PopBlends system. 
Participants were first asked to pick one pop culture domain that interested them most, from the set we provided: Star Wars, Friends, Harry Potter, Game of Thrones and Breaking Bad. 
For that pop culture domain, they were then asked to make blend ideas for each of the following product topics: 1) toothbrush, 2) airplane, 3) pizza, 4) beer, 5) cell phone, 6) swimming.  
The pop culture domains and products were the same as used in the annotation study. 
They were selected based on the criteria of diversity and randomness (explained in Section \ref{topic_selection}), which is consistent with previous literature \cite{metamap, visiblends, solvent, iconate}. 
Participants alternated between the baseline and the system condition.
Participants were randomly assigned to a condition order (either baseline first and then system or system first and then baseline)
that was counter-balanced to prevent a learning effect. 

At the beginning of the experiment, participants were introduced to the basic concepts of pop culture blend design. 
Participants then ideated blends for the six pop culture-product pairs. 
For each pair they were given five minutes to ideate as many blends as they could. 
In both conditions, participants saved useful images and came up with ideas in a PowerPoint. 
Before they used PopBlends, they were given a short presentation about the system.
After ideating blends for a pair, participants completed a NASA-TLX \cite{hart2006nasa} questionnaire, to measure their perceived mental workload. 
After all six pop culture-product pairs, participants were asked a series of questions about their experience of using the system and their views on system outputs in a 10-minute semi-structured interview.  
Finally, they were asked to pick one of their favorite ideas and make a final pop culture blend image. 

\subsection{Results}
\subsubsection{With PopBlends, participants ideated significantly more blends than with the baseline.}
Participants came up with an average of 2.03 (SD=0.84) ideas for a pop culture-product pair with PopBlends, compared to 0.87 (SD=0.56) ideas with the baseline.
Since the study was within subjects and involved count data, we ran a paired-sample Wilcoxon test and found the difference between these means statistically significant (p < 0.001). 
For several difficult pop culture-product combinations, such as Game of Thrones and cell phone, users would sometimes not have any ideas for a blend with the baseline. 
However, with PopBlends, they were always able to come up with at least one idea for a blend. 
When asked which tool helped more with coming up with ideas, all 10 participants agreed that PopBlends helped more.
The three main reasons mentioned by users were (1) Internet search provided too specific results that did not help them explore around the keyword and it was easy to get stuck in thinking of other keywords for new ideas (P5, P6 and P9). 
(2) PopBlends expanded both the product and pop culture domains (P1, P3 and P8) and provided diverse connecting concepts (P2, P7 and P10). 
(3) Lots of the results of scene pairs under the connecting concept were meaningful and saved users’ time in further search of fitting scene pictures (P4 and P7).

\subsubsection{Using PopBlends required significantly less mental demand and led to higher perceived performance.}

\begin{table}[t]
\begin{tabular}{| c | c | c | c |}
\hline
 \textbf{} & PopBlends & Baseline & p-value \\
\hline
Mental Demand & 2.85 (1.08) & 5.80 (2.14)  & \textbf{0.004}  \\ \hline
Physical Demand & 2.40 (2.32) & 3.10 (2.88)  & 0.102  \\ \hline
Temporal Demand & 3.00 (2.39)  & 4.65 (2.47)  & 0.028  \\ \hline
Performance & 2.10 (0.91) & 3.60 (2.17) & \textbf{0.004}  \\ \hline
Effort & 3.05 (1.50) & 5.00 (1.84) & \textbf{0.006} \\ \hline
Frustration & 1.65 (1.60) & 3.25 (2.75) & 0.026  \\ \hline
\end{tabular}
\caption{NASA-TLX questionnaire results comparing PopBlends and Baseline, where means, standard deviations (in parentheses) and p-values for the 6 paired-sample Wilcoxon tests with Bonferroni correction are reported. Bolded p-values are statistically significant. }
\label{table:tlx}
\end{table}

Results of the NASA-TLX questionnaire are shown in Table \ref{table:tlx}.
Participants found ideating blends with PopBlends significantly less mentally demanding than with the baseline, rating mental demand on average 2.85 with PopBlends and 5.80 with the baseline ($V$ = 45, $p$ = 0.004).
With the baseline, users had to come up with search keywords themselves and had trouble thinking of the connecting concept to blend the two domains.
Meanwhile, with PopBlends, lots of calculations were already done by the system to show users relevant keywords and possible candidate connecting concepts. 
Aligning with this result, perceived effort ($V$ = 44, $p$ = 0.006) was also significantly lower for PopBlends compared to the baseline. 
When using the baseline, users often needed to deal with more non-domain related information if the keyword was not carefully considered (e.g., when users searched ``friends + cell phone'', many pictures of friends chatting happily appeared, but none of them were related to the TV show Friends). 
Meanwhile, users reported they performed significantly better with PopBlends than with the baseline, with an average value of performance of 2.10 with PopBlends and 3.60 with the baseline ($V$ = 45, $p$ = 0.004). 
By browsing the result pairs PopBlends provided, users more successfully found inspirations and materials they needed to do the pop culture blend tasks.

\subsubsection{Users’ feedback on PopBlends system design} 
Generally, users appreciated that PopBlends specifically presents the connecting concepts that bridge the pop culture and product domains, as \textit{``the connecting concept provides a good hint to find the basis for blending two different domains, which was a challenge when I was using the Internet search''}, said P1. 
Under each connecting concept, PopBlends shows relevant pop culture and product scene images. 
\textit{``This is very handy, and I can often find useful images to use. It would be nicer if I could know how relevant these scenes are to the connecting concept, because sometimes I feel that the scenes listed, especially the pop culture scenes, are not as relevant as I thought they would be.'' }, commented by P9. 
In addition, half of the users mentioned that they liked the word selection feature for the product SWOW words because it offered a new possibility when they were stuck on the current result. 
For example, when P7 was trying to blend Friends and airplane, \textit{``the results for `airplane' were good and relevant, but I was getting a little tired of making connections from the `airport' and `travel' dimensions. 
So I tried adding several different related words to see the updated results, and I found a few I liked, especially the `airplane + seat' leading to Joey and Chan's recliner idea''}.

\begin{figure*}
\centering
\includegraphics[width=0.68\textwidth]{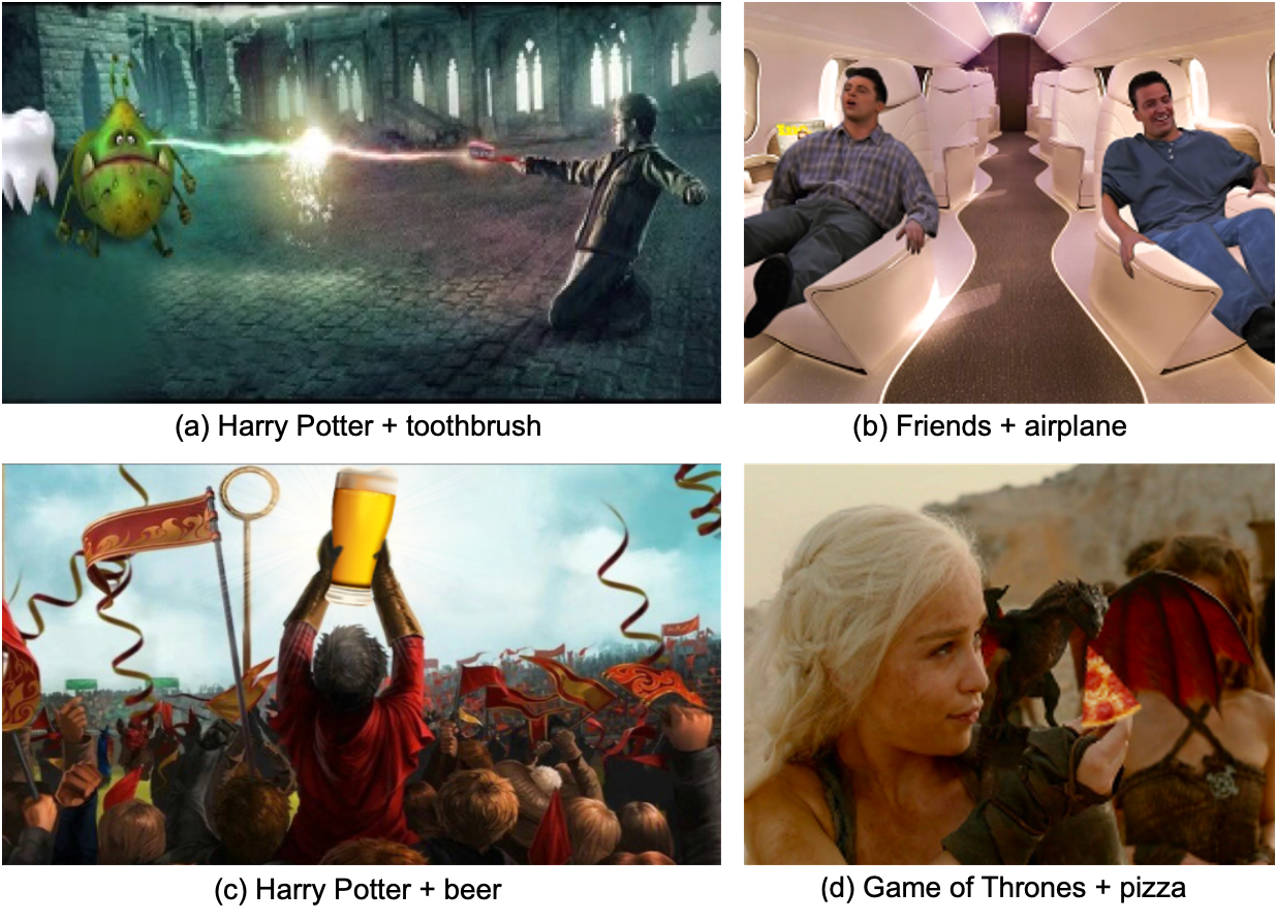}
\caption{Pop culture blends created by participants using PopBlends, with taglines of (a) Kill the evil plaque with your toothbrush wand (b) Seats of our airplane are so relaxing and comfortable as Joey and Chan’s (c) Yeah we finally win this Quidditch beer cup (d) Reward your baby dragon with pizza after a fight.}
\label{fig:evaluation_showcase}
\end{figure*}

\subsubsection{Users’ feedback on PopBlends results}
Overall, users commented that the three types of results (No-GPT, Half-GPT, Full-GPT) offer different but equally inspiring directions to blend.

No-GPT results, according to the users, were relevant and useful because the connecting concept always came back to a specific element in the plot sentences.
For example, P4 mentioned that the No-GPT approach succeeded in finding diverse connections for pizza in Harry Potter movies, commenting \textit{``When I searched Harry Potter and pizza on the Internet, I found almost nothing useful, but PopBlends was able to automatically generalize the keyword from pizza to food and show multiple plots related to food in Harry Potter.''} 
However, sometimes the plot matching method is limited because there may be no product-relevant elements in the plots.
For example, P10 (used PopBlends for Game of Thrones and toothbrush, pizza, cell phone) mentioned the No-GPT connecting concepts were \textit{``very relevant to Game of Thrones, but not that related to the product''}. 
This was the case because the pop culture wiki plot collection sometimes did not have words that matched well with the product.
However, even when the plot sentence did not perfectly match the product concept, it can still sometimes inspire users.
For example, P7 commented, \textit{``I love the matching of swimming to Friends' plot of Chandler relaxing in the bath. It was great! I was kind of stuck in finding people swimming fast in the pool scene in Friends, but they can relax in the pool as well.''}

Half-GPT results overcame the limitation of relying only on plot sentences to find connections, and over half of the users mentioned that matching from product concepts to entity attributes was beneficial to their divergent thinking.
As P9 explained, \textit{``Half-GPT is good, so instead of being stuck looking for some very precise plot connection, I can think of something related to the featured pop culture entity.''}
For example, when connecting Star Wars to pizza, Half-GPT found that exclaiming ``yippee'' (an Ewok's trait) was a matching attribute of pizza, and P9 generated a creative ad for a pizza party at the Ewoks', based on this prompt. 
However, due to the limited correctness of GPT-3 responses on the attributes of pop culture entities, three users mentioned that sometimes the Half-GPT results did not make much sense.

Full-GPT results provided the most diverse connections that users sometimes found very surprising and novel.
\textit{``It was surprising that PopBlends associated beer with Quidditch cup because it can `be used as a prize'}, as P2 mentioned, \textit{``I don't think I can ever come up with this myself.''} 
Unrestricted by the activities, adjectives, and catchphrases of the collected pop culture entities, Full-GPT provides a variety of connections that while not always true in the pop culture realm, were often imaginative.
Some users also noted that the Full-GPT method even finds shape connections between pop culture objects and products, such as P9's reference to PopBlends linking toothbrushes to lightsabers on the basis that they both have ``long and thin handles''. 
However, according to users, the quality of the Full-GPT connections varied the most among the three types of results, sometimes leaving them frustrated and not understanding why. 
For example, GPT-3 connects swimming to Star Wars based on ``Anakin Skywalker's death in water''. 
\textit{``But seems that never happened in Star Wars''}, P1 said. 
In general, however, most users appreciated the novelty of the Full-GPT results and the support for their ideation. 

\subsubsection{Showcase of pop culture blends created with PopBlends} 
Figure \ref{fig:evaluation_showcase} shows four examples of pop culture blends made by users using PopBlends. 
Example (a) is from P4, inserts toothbrush into Harry Potter fighting Voldemort scene in place of Harry's wand, inspired by No-GPT result: ``Harry tries the expelliarmus charm to block Voldemort's attempted killing curse'' with the connecting concept of ``curse''.
Example (b) is from P7, inserts relaxing Joey and Chandler into the passengers sitting comfortably in airplane seats scene in place of the passengers. 
P7 searched ``airplane'' and added ``seat'' from SWOW words, and was inspired by the No-GPT result of ``When Rachel goes out with Joey to buy a new recliner…'' plot with the connecting concept of ``sit''. 
P7 also used the airplane plus seat scene of ``sitting comfortably in the first class''.
Example (c) is from P2, inserts beer into the Harry Potter winning Quidditch game scene in place of the Quidditch cup.
P2 is inspired by Full-GPT result of connecting ``beer'' to Quidditch cup with the concept of ``use as a prize''. 
Example (d) is from P10, inserts pizza into Daenerys feeding her dragon scene in place of the meat, inspired by Half-GPT result of connecting ``pizza'' to Daenerys (mother of dragons) with the concept of her activity ``feed dragons''. 
\section{Discussion}
\subsection{Do Machines Need Divergent and Convergent ``Thinking''?}
Divergent and convergent thinking is a way of structuring creativity that has been shown to improve results for people.
However, we should consider whether this approach can also improve GPT-3's ability to be creative. 
In the Full-GPT strategy, we asked GPT-3 to find connecting concepts without explicit divergent and convergent steps.
It was able to come up with many useful outputs. 
However, this approach has limitations.
The Half-GPT approach (which does use explicit divergent and convergent steps) was able to come up with additional connections and with different characteristics. 
For example, across all our data, Full-GPT never proposed a connection based on a character's catchphrase, but it was a useful type of connection for many successful blends. 
This is consistent with the findings from the design literature on people showing that most people can ideate a small number ideas on their own \cite{paulus1993perception}, but to get beyond that number, it is useful to have different brainstorming strategies. 

The ideation literature also advises that the best way to have a good idea is to have a lot of ideas. 
Thus, there is a divergent phase of producing as many ideas as possible, then a convergent phase of selecting the best ones. 
Since all three GPT-3 strategies are good, but noisy (about 50\% accurate), we argue they are all useful to try---particularly since they produce different types of connections. 
Whether a creative task is done by a human or a machine, having an explicit creative process (such as divergent and convergent steps) allows us to edit intermediate responses, fix localized problems in the process, and construct a tighter feedback loop. 
Although we can try to jump straight to the answer, it is good to have a process to fall back on.

\subsection{Combining Language Models with Knowledge Bases} 
Knowledge bases and language models are complementary sources of information.
While knowledge bases are typically quite accurate and well-structured, they are limited in scope based on the information they can scrape. 
On the other hand, large language models have access to vast amounts of data, but are unstructured and can be inaccurate due to their propensity to hallucinate.  Thus, there is an opportunity to combine these approaches to get the best of both: more data with higher accuracy.
Our Half-GPT approach did this---it used a structured knowledge base to provide a list of characters, organizations, locations and objects, and then used GPT-3 to extend them. 
This enabled us to populate adjectives, activities and catchphrases for every entity in the knowledge base. 
By combining the structured nature of the knowledge base with the vast amount of data from GPT-3, we were able to get additional information that covered all the important entities.

Although GPT-3 has already been publicized as a way to construct knowledge bases without scraping \cite{gpt_as_kb}, we find that the information in GPT-3 is quite different and complementary to the information in a knowledge base.
Knowledge bases tend to focus on indisputable facts such as listing all the characters in Friends. 
Whereas GPT-3 can produce this type of facts, it can also produce associative data that is not strictly factual such as what traits we associate with characters. (Most people associate Joey in Friends with being dim and Monica with being obsessive and controlling.) 
These are not the typical facts stored in knowledge bases because they are more subjective and difficult to scrape. However, they are incredibly useful for creative tasks, and GPT-3 can provide this complementary information.

\subsection{Potential for Ensemble Learning}
\label{ensemble_learning}
One of our main findings was that No-GPT, Half-GPT and Full-GPT were all equally accurate but with different characteristics. 
A consequence of this is we probably should use them all. 
This is consistent with ensemble learning models \cite{sagi2018ensemble}, where each method is a ``weak learner'' that cannot generalize to all instances in the data, but a combination of multiple methods can compensate for each other's errors and exceed the performance of a single method. 

A challenge in applying ensemble approaches is to determine when each method in the ensemble is most accurate for which inputs. 
Unfortunately, we found that GPT-3 lacks this type of accuracy information. It generally does not have ``awareness of restrictions'' \cite{razniewski2021language}. 
In our prompting experiments, GPT-3 rarely answered ``I don't know'' or ``It doesn't exist'', and it almost always gave an answer even if the question was ridiculous. 
For example, when prompted with the questions like ``What are the catchphrases of Dementors in Harry Potter?'' (an entity that is mute), it provides many false answers such as ``I'm coming for you'', ``You're mine'', and ``You're next''. 
In the future, if GPT-3 can be imbued with a better awareness of its restrictions, the potential for combining multiple GPT-based generation methods into ensemble approaches could be increased.
\section{Limitations and Future Work}
Currently, the pop culture domains that PopBlends focuses on are all movies and television series. 
Many of the pop culture blends on social media also make reference to movies and television series, probably because they are highly visual (unlike music) and pervade culture. 
However, pop culture encompasses many more things---music, toys, politics, sports, technology, etc. 
Our method of blending currently requires plot summaries, which only exist for film and TV entertainment.
Extending to other pop culture domains would require finding a new way of finding scenes. 
There is a possibility that GPT could replace the role of the plot summary. 
GPT could almost certainly identify the major people, places and things in a domain. 
It might also be able to list scenes, as we discussed in Section \ref{diff_gen_ways}---but it is unclear if those scenes would have enough fact-based information, as opposed to hallucinations. 
This is an approach we intend to try in the future.

PopBlends focuses on suggesting images that could be blended, but it does not actually blend them. 
Full automatic blending may or may not be possible, but it is an obvious next goal. 
Text-to-image models like Dall-E \cite{ramesh2021zero} bring the possibility of making final blends based directly on the textual description of the idea. 
However, in our early experiments, Dall-E did not work as well as expected, in particular it did not seem to correctly depict domain-specific objects like TIE fighters, which made the blends lacking in recognizability. 
Even if full automation is not possible, simple machine vision may be able to find replacing objects from the images based on the text descriptions from the connecting words. 
Even a low quality blend may be acceptable to post online, as much of meme culture is made up of low-quality image modification. 
Other systems have shown how an additional round of divergent and convergent thinking can improve image blends ~\cite{VisiFit}. 
Perhaps those approaches could be applied here. 

According to our study, GPT-based results proved to be helpful to users, but further improvements can be made.
In the future, we can do more GPT prompt engineering, such as providing examples in the prompts, to make GPT-based results even more accurate. 
Additionally, we could go beyond fully automatic suggestions and add more user interaction.
For example, we can allow users to call GPT-3 multiple times or provide more contexts when needed. 

PopBlends is designed to help amateurs as many creativity support systems do  \cite{visiblends, symbolfinder, metamap, solvent}. 
Amateurs are important to support in creative tasks because not everyone can spend years in design school to learn creativity approaches. 
In the future work, we would like to explore how experts might benefit as well.
The creativity systems mentioned above have explored this and found they benefit experts as well as amateurs.
This is likely because divergent and convergent thinking is mentally demanding due to fundamental cognitive and memory limitations of the human brain, regardless of the level of expertise. 
We would like to release the tool to the wild for further exploration, for example, to see what pop culture and product domains users pick, how useful they find the results, and whether they want the system to be more interactive.
\section{Conclusion}
This paper presents PopBlends, a system that automatically finds connecting concepts between inputs and suggests images the user could pair in a pop culture blend. 
The system uses two rounds of divergent and convergent processes based on structured NLP techniques and large language models. 
A ten-participant user study shows that the system helped people come up more ideas, and with significantly less mental demand compared to the baseline condition. 
We discuss how divergent and convergent thinking is useful to structure not only human creativity, but also machine creativity.

\balance

\bibliographystyle{ACM-Reference-Format}
\bibliography{sample-base}

\clearpage
\appendix
\begin{minipage}{\textwidth}
\section{Connecting Concepts of Airplane and Pop Culture Domains}
\setcounter{table}{0}
\renewcommand{\thetable}{A\arabic{table}}

\begin{table}[H]
\footnotesize
\begin{tabular}{lp{4.3cm}p{3.6cm}p{4.1cm}}
\toprule
\textbf{Domains} & \textbf{No-GPT} & \textbf{Half-GPT} & \textbf{Full-GPT} \\
\midrule
Star Wars & \textit{Falcon} \newline{(Before the \textit{Falcon} can reach Alderaan...)} & \textit{piloting aircraft} \newline{(activity of Bespin)}   & \textit{his breathing apparatus} \newline{(associated with Darth Vader)} \\
\midrule
Friends & \textit{airport} \newline{(Monica persuades Ross...to the \textit{airport}...)} & \textit{``I got off the plane''} \newline{(catchphrase of Rachel)}   & \textit{his fear of flying} \newline{(associated with Chandler)}  \\
\midrule
Harry Potter & \textit{flying} \newline{(Ron…rescued Harry in \textit{flying} Ford Anglia)} & \textit{flying on broomsticks} \newline{(activity of Harry  Potter)} & \textit{both fast and have wings} \newline{(associated with Golden Snitch)}  \\ 
\midrule
Game of Thrones & \textit{flying} \newline{(Tyrion sees...dragon \textit{flying} overhead)} & \textit{flying on dragons} \newline{(activity of Daenerys Targaryen)} & \textit{his travel from King's landing to Pentos }\newline{(associated with Tyrion Lannister)} \\ 
\midrule
Breaking Bad & \textit{crew} \newline{(Walter surrenders...Jack's \textit{crew} arrive)} & \textit{``I am the danger''} \newline{(catchphrase of Walter White)} & \textit{his witness of an airplane crash} \newline{(associated with Walter's swimming pool)} \\
\bottomrule
\end{tabular}
\caption{Example connecting concepts of airplane and pop culture domains across No-GPT, Half-GPT and Full-GPT strategies. Note that ``crew'' has different meanings in the domain of Breaking Bad and airplane, ``I am the danger'' is Walter White's catchphrase but not fully connected to airplane, Chandler was afraid to take the flight to Yemen but mainly because of Janice.}
\label{tab:airplane_examples}
\end{table}

\section{Pop Culture Entities}
\setcounter{table}{0}
\renewcommand{\thetable}{B\arabic{table}}

\begin{table}[H]
\footnotesize
\begin{tabular}{p{1.8cm}p{3.2cm}p{3cm}p{3cm}p{3cm}}
\toprule
\textbf{Domains} & \textbf{Characters} & \textbf{Organizations} & \textbf{Locations} &\textbf{Objects / Others} \\
\midrule
Star Wars  
& Luke Skywalker, Darth Vader, Han Solo, Leia Organa, Obi-Wan Kenobi, Chewbacca, R2-D2, Lando Calrissian, Jedi knights, Jabba the Hutt 
& Force, Rebels, \textit{Millennium Falcon}, Rebel Alliance, Rebel Fleet, \textit{TIE}, Ewoks, \textit{X-wing}, Wookiee, \textit{Admiral Ackbar} 
& Death Star, Alderaan, Endor, Tatooine, Dagobah, Hoth, Mos Eisley, Bespin, Cloud City, \textit{Galactic Republic}
& lightsaber, Imperial fleet, planet, Emperor, shield, father, Galactic Empire, dark side, city, base
\\
\midrule
Friends 
& Ross, Rachel, Joey, Chandler, Monica, Phoebe, Mike, Emily, Ben, Pete Becker
& friends, Bloomingdale's, Gucci, Discovery Channel, Knicks, Animal Control, Soap Opera Digest, Rangers, Law \& Order, the Post
& London, Tulsa, Vail, Las Vegas, China, New York City, Paris, Broadway, Vermont, Greece
& Days of our Lives, relationship, Valentine's day, C.H.E.E.S.E, FICA, the Shining, Little Women, Baby Got Back, couple, Wonderful Tonight
\\
\midrule
Harry Potter 
& Harry Potter, Ron Weasley, Hermione Granger, Sirius Black, Draco Malfoy, Lord Voldemort, Dobby, Severus Snape, Lucius Malfoy, Ginny Weasley 
& Death Eaters, Dursleys, Dementors, Defence Against the Dark Arts, Ministry of Magic, Order of the Phoenix, Blacks, Potters, Weasleys, Snatchers
& Hogwarts, Azkaban, Room of Requirement, Gringotts, Whomping Willow, Burrow, Forbidden Forest, Durmstrang, Diagon Alley, Great Hall
& Hogwarts Express, trio, Horcrux, locket, Killing Curse, Patronus Charm, father, Sword of Gryffindor, Quidditch World Cup, school
\\ 
\midrule
Game of Thrones 
& Jon Snow, Daenerys Targaryen, Tyrion Lannister, Sansa Stark, Arya Stark, Jaime Lannister, Cersei, Theon Greyjoy, Jorah Mormont, Melisandre
& Wildlings, Unsullied, White Walkers, slaver, Brotherhood, Dothraki, Lannisters, Sons of the Harpy, Kingsguard, watchmen
& King's Landing, Winterfell, Meereen, Castle Black, Dragonstone, Braavos, Yunkai, Dorne, Essos, Keep
& Iron Throne, dragons, murder, soldiers, Dragonglass, \textit{Faith Militant}, royal wedding, death, Needle, \textit{Second Sons}
\\ 
\midrule
Breaking Bad 
& Walter White, Jesse Pinkman, Hank Schrader, Skyler, Gus Fring, Jane Margolis, Mike, Walter Jr., Ted, Saul Goodman
& IRS, police, \textit{ATM}, Aryan brotherhood, Los Pollos Hermanos, Gray Matter, DEA agent, Mexicans, Denny's, ASAC
& Albuquerque, Mexico, El Paso, Czech Republic, Juarez, Tohajiilee Indian Reservation, Cayman Islands, Casa Tranquila, New Hampshire, Nebraska
& meth, lab, Leaves of Grass, money, dealers, RV, methylamine, product, business, phone
\\
\bottomrule
\end{tabular}
\caption{Pop culture entities collected by PopBlends in types of characters, organizations, locations and objects/others. Entities in italics are potentially assigned to the wrong category.}
\label{tab:pop_entities}
\end{table}

\end{minipage}

\end{document}